\documentclass[twocolumn]{aastex7}

\usepackage{amsmath}
\graphicspath{{./}{figures/}}

\shorttitle{Expectation for the Gamma-Ray Emission from PISN}
\shortauthors{Ryo Sawada}

\begin{document}


\title{Expectation for the Hard X-Ray and Gamma-Ray Emission from Pair-Instability Supernovae}

\author[orcid=0000-0003-4876-5996,gname=Ryo, sname='Sawada']{Ryo Sawada} 
\affiliation{Institute for Cosmic Ray Research, The University of Tokyo, Kashiwa, Chiba 277-8582, Japan}
\email[show]{ryo@g.ecc.u-tokyo.ac.jp}

\correspondingauthor{Ryo Sawada}

\begin{abstract}
Pair-instability supernovae (PISNe) are predicted thermonuclear explosions of massive stars with helium core masses exceeding $\sim 65M_\odot$ and synthesize substantial amounts of radioactive $\mathrm{^{56}Ni}$ ($M(\mathrm{^{56}Ni})\sim60M_\odot$ in extreme cases). 
To investigate their observational signatures, we developed a multi-D Monte Carlo radiation transport code, assuming spherical symmetry in the background medium and the photon sources distribution, and performed simulations of gamma-ray and hard X-ray emissions from the decay chain  $\mathrm{^{56}Ni}\to\mathrm{^{56}Co}\to\mathrm{^{56}Fe}$. 
We find that key gamma-ray lines (847 and 1238 keV) from $\mathrm{^{56}Co}$ decay in the $130M_\odot$ helium core model can be detected up to 300–400 Mpc by next-generation MeV gamma-ray telescopes. 
In contrast, the signals from the $100M_\odot$ model remain below the detection limits. 
Our results provide the template for gamma-ray follow-up observations of PISNe.
Considering theoretical predictions and observational constraints, we estimate PISN event rates within 300 Mpc to be approximately 0.01–0.1 events per year, highlighting their rarity but also emphasizing their feasibility as targets for future gamma-ray observations over the decade.
\end{abstract}

\section{Introduction}\label{sec:intro}
Pair-instability supernovae (PISNe) are theorized thermonuclear explosions marking the end of very massive stars \citep{1967PhRvL..18..379B,1967ApJ...148..803R}, and are proposed to occur when the helium core mass lies between about 65 and 135 $M_\odot$ \citep{2002ApJ...567..532H}. 
With predicted explosion energies reaching nearly $\sim 10^{53}$ erg, theoretical models suggest that PISNe can synthesize vast amounts of radioactive $^{56}$Ni--—up to $\approx 60M_\odot$\citep{2002ApJ...567..532H}---potentially driving exceptionally luminous optical light curves comparable to superluminous supernovae (SLSNe) \citep{2011ApJ...734..102K,2013MNRAS.428.3227D,2017A&A...599L...5G}. 

Following the observational discovery of SLSNe, which exceed the luminosities of ordinary supernovae by more than an order of magnitude \citep[see][for a recent review]{2019ARA&A..57..305G}, initial interpretations linked them theoretically to PISNe \citep[][]{2007ApJ...666.1116S,2009Natur.462..624G}. 
However, subsequent detailed observational studies revealed discrepancies between the light curves and spectral features predicted by PISN models and those observed in the majority of SLSNe, implying alternative powering mechanisms for most of these extraordinary events \citep{2012MNRAS.426L..76D,2013ARA&A..51..457N,2016MNRAS.455.3207J,2017MNRAS.468.4642I,2018SSRv..214...59M,2019ApJ...882...36M}. 
Thus, despite several promising candidate events \citep{2016ApJ...831..144L,2017NatAs...1..713T,2018MNRAS.479.3106K,2024A&A...683A.223S} and potential indirect signatures \citep[][]{2014A&A...572A..80A,2022ApJ...925..111I}, no definitive observational confirmation of a PISN has yet been achieved.
Confirming the existence of PISNe remains crucial, as it would not only validate fundamental predictions of stellar evolution theory \citep{2012ARA&A..50..107L} but also provide deeper insights into the chemical evolution of the universe \citep[][]{2013ARA&A..51..457N,2025arXiv250311457N} and constrain black hole mass distributions relevant to gravitational-wave astronomy \citep[][]{2016A&A...588A..50M,2019ApJ...887...53F}.

Given the significant quantity of radioactive $^{56}$Ni synthesized during PISNe explosions \citep[$M_\mathrm{^{56}Ni}\lesssim 60M_\odot$;][]{2002ApJ...567..532H,2024MNRAS.531.2786K}, $\gamma$-ray and hard X-ray observations offer a powerful observational tool. 
The radioactive decay chain of $\mathrm{^{56}Ni}\to\mathrm{^{56}Co}\to\mathrm{^{56}Fe}$ generates characteristic gamma-ray lines around $\sim1$ MeV.
Most of these emitted gamma rays provide the energy source for the optical light curve of the supernova \citep[e.g.,][]{1982ApJ...253..785A}, and some of them would reach to us as gamma rays or hard X-rays \citep[e.g.,][]{2006ApJ...644..385M}. 
As a simple approximation, the peak gamma line fluxes occur near a time, $t_\mathrm{peak}\propto(\kappa_\gamma(M_\mathrm{ej}^2\tau/E))^{1/3}$, with a peak flux $F_\mathrm{peak}\propto M(\mathrm{^{56}Ni})\cdot e^{-(t_\mathrm{peak}/\tau)}/\tau$, where $M_\mathrm{ej}$ is the ejecta mass, $E$ is the explosion kinetic energy, $\tau$ is the mean life of the decay, and $\kappa_\gamma$ is the Compton opacity of the line. 
Given that SN 1987A produced approximately 0.07 $M_\odot$ of $^{56}$Ni \citep[][]{1989ARA&A..27..629A} and typical Type Ia supernovae produce around 0.6 $M_\odot$ \citep[][]{1984ApJ...286..644N}, PISNe—with their dramatically larger $^{56}$Ni yields of up to 60 $M_\odot$—could potentially be detectable at distances more than an order of magnitude greater than typical detection limits for Type Ia supernovae \citep[$\sim 10$ Mpc;][]{2014ApJ...786..141T}.

While significant progress has been made in understanding the optical light curves of PISNe \citep[e.g.,][]{2011ApJ...734..102K,2013MNRAS.428.3227D}, direct observational signatures in the gamma-ray and hard X-ray regimes have been less explored theoretically in detail. 
Some studies, such as those by \cite{2015MNRAS.454.4357K} and \cite{2017MNRAS.464.2854K}, have investigated aspects of PISN observables by consistently modeling their hydrodynamic evolution and subsequent light curves. 
These models provide crucial insights into the overall energy budget and light curve morphology, including contributions from radioactive decay. 
However, while their work included gamma-ray energy deposition as part of the light curve modeling, detailed predictions of the emergent gamma-ray spectra and their detectability by next-generation MeV gamma-ray telescopes were not the primary focus.

Historically, gamma-ray emission from supernovae has been rarely observed due to their intrinsic faintness and the limited sensitivity of past gamma-ray missions. 
SN 1987A remains the only core-collapse supernova with clear gamma-ray observations \citep[][]{1987Natur.330..227S}. 
Similarly, SN 2014J remains the only Type Ia supernova directly observed in gamma rays, with emission lines from $^{56}$Co detected approximately 10 days after the explosion \citep[][]{2015A&A...574A..72D}. 
Next-generation gamma-ray telescopes, including e-ASTROGRAM \citep[from 0.3 MeV to 3 GeV;][]{2017ExA....44...25D}, AMEGO-X \citep[from 100 keV to 1 GeV;][]{2022icrc.confE.649F}, and COSI mission \citep[0.2–5 MeV;][]{2022icrc.confE.652T}, are expected to greatly enhance sensitivity and spectral resolution, significantly improving the prospects for detecting gamma-ray signatures from PISNe.

In this paper, we present the first comprehensive theoretical calculations of gamma-ray and hard X-ray emission originating from radioactive $^{56}$Ni synthesized in PISNe ejecta. 
This paper is organized as follows.
In Section \ref{sec:model}, we briefly summarize the input PISN models, and details of the computational method are presented in Section \ref{sec:method}.
In Section \ref{sec:spctra}, we show the overall synthetic spectra. 
Section \ref{sec:lc} focuses on light curves of 847 keV and 1238 keV lines, and Section \ref{sec:discus}
we discuss the event rates of PISNe.
Finally, in Section \ref{sec:concle} conclusions and discussion are presented.

\section{Model}\label{sec:model}
We use PISN explosion models based on 1D simulations computed using the {\tt MESA-r24.08} software instrument for stellar evolution \citep{2011ApJS..192....3P,2013ApJS..208....4P,2015ApJS..220...15P,2018ApJS..234...34P,2019ApJS..243...10P}.
We compute a set of nonrotating models of helium stars at a metallicity of $Z=10^{-5}$ as the protosolar abundance reported by \cite{2009ARA&A..47..481A}. 
The initial models were helium stars with masses of $100$ and $130M_\odot$, which were chosen after confirming that pair-instability explosions would occur \citep[see also][]{2017ApJ...836..244W,2019ApJ...887...53F}.
For the configuration of physical parameters, we rely largely on the default setting of the {\tt ppisn} test suite for {\tt MESA-r24.08}, as described in \cite{2019ApJ...882...36M}.
We use the {\tt mesa128.net} nuclear network within MESA to track 128 isotopes and nuclear burning processes, particularly to estimate the amount of produced in explosive events.

\begin{figure}[htpb]
\centering
  \includegraphics[width=0.45\textwidth]{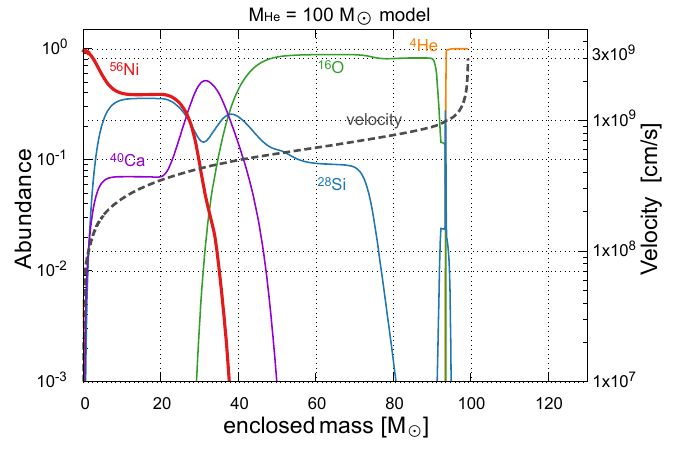}
  \includegraphics[width=0.45\textwidth]{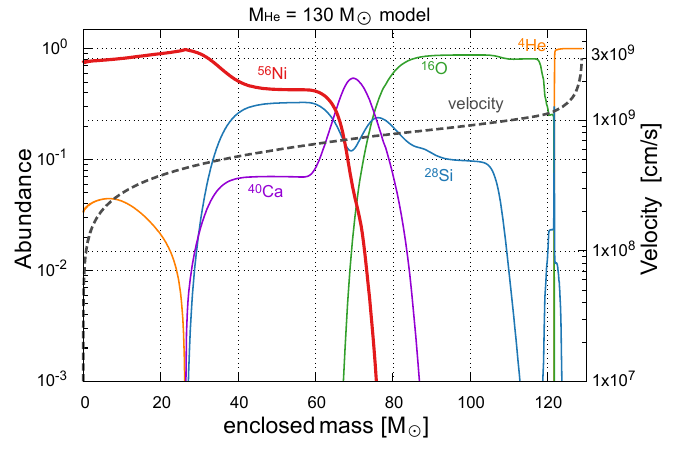}
    \caption{
    Distribution of major nuclear species and expansion velocity as a function of enclosed mass coordinate at the homologous expansion phases, for different PISN models with helium core mass $M_\mathrm{He}=100\,M_\odot$ (top) and $M_\mathrm{He}=130\,M_\odot$ (bottom).
    The left axis shows the mass fractions of 56Ni and other key isotopes, while the right axis indicates the local expansion velocity.}
\label{fig:abundance}
\end{figure}

Figure \ref{fig:abundance} show results of the explosive nucleosynthesis calculation for the different PISN models with helium core mass $M_\mathrm{He}=100\,M_\odot$ and $M_\mathrm{He}=130\,M_\odot$, respectively.
In the $M_\mathrm{He}=100\,M_\odot$ case, about $13M_\odot$ of $\mathrm{^{56}Ni}$ are synthesized and the explosion energy are $4.5\times10^{52}$ ergs. 
In the $M_\mathrm{He}=130\,M_\odot$ case, about $44M_\odot$ of $\mathrm{^{56}Ni}$ are synthesized and the explosion energy are $9.5\times10^{52}$ ergs.

One crucial nuclear reaction that can significantly affect PISN explosion conditions and the synthesized mass of $\mathrm{^{56}Ni}$ is $\mathrm{^{12}C(\alpha,\gamma)^{16}O}$ reaction rate \citep[][]{2019ApJ...887...53F,2020ApJ...902L..36F,2024MNRAS.531.2786K}.
However, in this study, we do not delve into the impact of its uncertainties on the explosion outcome. 
Instead, to maintain consistency with the standard rates adopted by \cite{2019ApJ...887...53F,2020ApJ...902L..36F} and \cite{2024MNRAS.531.2786K}, we employ a modified $\mathrm{^{12}C(\alpha,\gamma)^{16}O}$ reaction rate taken from \cite{2002ApJ...567..643K}. 
This choice ensures alignment with previous studies, in which variations in this rate were shown to alter the threshold for PISN explosions as well as the $\mathrm{^{56}Ni}$ yield.

\section{Method}\label{sec:method}
We developed a Monte Carlo (MC) code that fully tracks the motion of photons in three-dimensional velocity space, with an assumed spherical symmetry in the background medium and the photon sources distribution, in order to solve the gamma-ray photon radiative transfer in supernova ejecta.
Our approach is based on the key methodologies from \citet{2019ApJ...882...22A} to model the propagation and interactions of photons produced by the decay of radioactive isotopes.

\begin{table}
\begin{center}
    \caption{List of the Radioactive Nuclei and Their Important Decay Lines in the Study} 
    \label{tbl:iso}
    \begin{tabular}{l|cc} \hline \hline
    Isotope & $E_\mathrm{\gamma‐ray}$ & Intensity $I_\mathrm{\gamma}$ \\
    & (keV) & ($\gamma$ decay$^{-1}$) \\  \hline 
    $^{56}$Ni $\rightarrow$ $^{56}$Co &158.38 keV & 0.988\\    
    $t_{1/2,\mathrm{rest}}=6.075$ days &269.50 keV & 0.365 \\
    &480.44 keV & 0.365 \\
    &749.95 keV & 0.495 \\
    &811.85 keV & 0.860 \\
    &1561.80 keV & 0.140\\ \hline 
    $^{56}$Co $\rightarrow$ $^{56}$Fe & 846.8 keV & 0.999 \\
    $t_{1/2,\mathrm{rest}}=77.24$ days & 1037.8 keV & 0.141\\
    &1238.29 keV& 0.665\\
    &1771.36 keV& 0.154\\
    &2598.50 keV& 0.170\\ \hline\hline
  \end{tabular}
\end{center}
\end{table}

\subsection{Monte Carlo Treatment of Photon Generation}
Gamma-ray photons are generated from the decay chains of $^{56}$Ni $\rightarrow$ $^{56}$Co $\rightarrow$ $^{56}$Fe, and important nuclear decay lines included in the code are listed in Table \ref{tbl:iso}.
The initial spatial position of the photon is determined by the spatial distribution of the parent nucleus; that is, the generated photon number at a given location is proportional to the local $^{56}$Ni density.
The direction of the photon is assigned isotropically in the rest frame of the expanding nucleus, and its energy is also initialized in that frame based on the energy of the radioactive decay. 
The dynamics of the ejecta are assumed to be homologous and spherically symmetric, which is expected to be a good approximation throughout the simulation timescales (see Section \ref{subsec:bm} in detail).
Each photon is converted to the rest frame using the Lorentz transformation, and this accounts for Doppler shifts and beaming effects due to homologous expansion.

The Lorentz time dilation factor is applied to the nucleus's half-life to account for radioactive decay in the relativistic expansion ejecta. 
Specifically, if $t_{1/2,\mathrm{rest}}$ denotes the half-life measured in the nucleus's rest frame, then observers in the lab frame see it extended to
\begin{equation}
    t_{1/2,\mathrm{lab}}= \gamma(v)\cdot t_{1/2,\mathrm{rest}}
\end{equation}
where $\gamma(v)=\sqrt{1-(v/c)^2}$ is the Lorentz factor associated with the local expansion velocity {of the ejecta $v$. 
Equivalently, the decay constant becomes $\lambda_\mathrm{lab}=\lambda_\mathrm{rest}/\gamma(v)$. 
We implement this correction on a shell-by-shell basis: for each radial shell moving at speed $v(r)$, the observed decay rate is reduced by the factor $\gamma(r)$. 
This ensures that faster-moving material experiences slower nuclear decay from the perspective of a distant observer, consistent with special relativity.

\subsection{PISN Ejecta as a Background Medium}\label{subsec:bm}
We assume that the dynamics of the PISN ejecta are homologous (i.e., velocity proportional to radius, $v \propto r$) and spherically symmetric throughout the simulation period.
This approximation is supported by our hydrodynamical models: the ejecta rapidly settle into a homologous expansion within approximately one day after shock breakout.
This behavior arises from the compactness of the progenitor star ($R_\mathrm{He-star} \approx 10^{12}$ cm) and the extremely high explosion energy, which drives shock velocities on the order of $v_\mathrm{sh} \approx 0.1c$.
We use the resulting ejecta profiles at this stage as the static background structure for our gamma-ray transport simulations.

We note that our hydrodynamic models do not include any local reheating or overpressure in the $^{56}$Ni-rich interior (the so-called Ni-bubble effect). 
In reality, energy deposited by $^{56}$Ni (and subsequent $^{56}$Co) decay creates an overpressure that inflates Ni-rich regions into `Ni bubbles' \citep[e.g.,][]{1998ApJ...496..946K}. 
As long as the blobs are opaque to gamma-rays, this effect can lead to lower densities, and hence lower opacities, which probably affects the spectra, or at least the timing of gamma-ray escape.

Furthermore, previous multidimensional studies have shown that Rayleigh–Taylor (RT) instabilities in PISNe cause only minor mixing of $\mathrm{^{56}Ni}$ \citep[][]{2017ApJ...846..100G}, with limited disruption of the large-scale spherical structure.
These findings support the validity of our assumption that deviations from strict spherical symmetry and homology have a negligible impact on the gamma-ray and hard X-ray emission modeled in this work.

\subsection{Photon Propagation and Interaction Physics}
As photons propagate through the ejecta, we consider interactions including photoelectric absorption, Compton scattering, and pair production. 
Bremsstrahlung and fluorescence emissions are neglected, because their contributions are at energies below the sharp photo-absorption cutoff \citep[][]{1991ApJ...375..221C,1995ApJ...455..215B}.

The photon path is traced until the accumulated optical depth $\tau$ reaches a threshold value, $\tau_{\mathrm{lim}}$, sampled from the exponential distribution:
\begin{equation}
\tau_{\mathrm{lim}} = -\ln \xi_1,
\end{equation}
where $\xi$ is a random number uniformly distributed in $[0,1]$. 
The optical depth $\tau$ along the photon path $L$ is calculated as follows: 
\begin{equation}
\tau = \tau_{\mathrm{CS}} + \tau_{\mathrm{ph}} + \tau_{\mathrm{pp}},
\end{equation}
where $\tau_{\mathrm{CS}}$, $\tau_{\mathrm{ph}}$, and $\tau_{\mathrm{pp}}$ represent the depths due to Compton scattering, photo absorption, and pair production, respectively. 
These are expressed as:
\begin{align}
\tau_{\mathrm{CS}} &= \int_L \sigma_{\mathrm{KN}}(r, E) n_e(r) \, dL, \\
\tau_{\mathrm{ph}} &= \int_L \sum_I \left(\sigma_{\mathrm{ph}}(r, E, I) n(r, I) \right) \, dL, \\
\tau_{\mathrm{pp}} &= \int_L \sum_I \left(\sigma_{\mathrm{pp}}(r, E, I) n(r, I) \right) \, dL,
\end{align}
where $n_e$ and $n(I)$ are the electron and element $I$ number densities, and the summation over $I$ includes all elements in the ejecta. 
For Compton scattering, the Klein-Nishina cross section $\sigma_{\mathrm{KN}}$ is used. 
For photoelectric absorption, cross sections $\sigma_{\mathrm{ph}}$ are taken from {\tt Xraylib} \citep{2004AcSpB..59.1725B,2011AcSpB..66..776S}.
For pair production, cross sections $\sigma_{\mathrm{pp}}$ are adopted from \cite{1995ApJ...446..766S}.

If an interaction takes place at the point $\tau=\tau_\mathrm{lim}$, the photon undergoes one of the three possible interactions.
The fate of the $\gamma$-ray packet is chosen randomly in proportion to the cross section of each possible interaction.
If the following condition is satisfied, the photon is absorbed:
\begin{equation}
\xi_2 \leq\cfrac{\tau_\mathrm{ph}}{\tau_\mathrm{cs}+\tau_\mathrm{ph}+\tau_\mathrm{pp}},
\end{equation}
where $\xi_2$ is also a random number uniformly distributed in $[0,1]$. 
In the simulation, we move on to the next MC photon from the current tracked photon.
Then, if the condition
\begin{equation}
\cfrac{\tau_\mathrm{ph}}{\tau_\mathrm{cs}+\tau_\mathrm{ph}+\tau_\mathrm{pp}}< \xi_2 \leq\cfrac{\tau_\mathrm{ph}+\tau_\mathrm{pp}}{\tau_\mathrm{cs}+\tau_\mathrm{ph}+\tau_\mathrm{pp}},
\end{equation}
is satisfied, the original photon is replaced by a pair of 511 keV photons with random directions in the locally comoving frame.

If neither photoelectric absorption nor pair production occurs, the photon undergoes Compton scattering. 
In our implementation, the scattering is handled in the locally comoving frame.
The photon's new energy and direction are determined according to the Klein–Nishina differential cross section \citep[e.g.,][]{1979rpa..book.....R}.
The energy shift is given by:
\begin{equation}
\frac{E_f'}{E_i'} = \frac{1}{1 + \frac{E_i'}{m_e c^2} (1 - \cos{ \alpha'})}~,\label{eq:scat_ene}
\end{equation}
where $E_f'$ and $E_i'$ are the initial and final photon energies in the comoving frame, and $\alpha'$ is the scattering angle in that frame.
The angular distribution is given by the Klein–Nishina formula:
\begin{equation}
\frac{d\sigma'_{\mathrm{KN}}}{d\Omega'} \propto \left( \frac{E_f'}{E_i'} \right)^2 \left( \frac{E_f'}{E_i'} + \frac{E_i'}{E_f'} - \sin^2 \alpha' \right)~,\label{eq:scat_angle}
\end{equation}
where $d\Omega'$ is the differential solid angle in the comoving frame.
We employ a rejection sampling technique, to sample new $\alpha$ from the distribution derived by substituting Equation \eqref{eq:scat_ene} into Equation \eqref{eq:scat_angle}.
After updating the photon's energy and direction, the next interaction is sampled, and this process is repeated until the photon either escapes from the ejecta or is absorbed.

\subsection{Numerical Setup and Some Assumptions}
The computational setup for this study is that the ejecta are mapped onto a radial grid of 1200 cells.
In our simulations, a photon packet for each nuclear decay line is assigned a uniform statistical weight.
To accurately represent the relative intensities of different decay lines, the number of photon packets generated for each line is scaled in proportion to its photon yield per decay, $I_\gamma$.
Specifically, we use $N_{\mathrm{packet,}\gamma}=I_\gamma\times N_\mathrm{packet, base}$, where the base packet number $N_\mathrm{packet, base}$ is $1.5\times10^7$ for spectral synthesis and $10^5$ for light curves calculations.
The integral calculation of optical depth $\tau$ along the photon path $L$ is discretized with $dL = 0.01 r$.

During photon propagation, the expansion of the ejecta is accounted for, with the local physical conditions updated based on the photon’s position and cumulative travel time.
However, we neglect the spread in photon escape times due to different scattering histories and assume that all photons generated at the same physical time are recorded as escaping simultaneously.
This approximation is justified because the maximum ejecta velocities $(\lesssim0.1c)$ are much smaller than the speed of light, and the ejecta become optically thin relatively quickly.
Consequently, flight time delays have a negligible impact on the emergent light curves and spectra \citep[see also][]{2006ApJ...644..385M}.

\section{$\gamma$- and Hard X-Ray Spectra}\label{sec:spctra}

Here, we present the time evolution and mass dependence of the synthetic $\gamma$- and hard X-ray spectra of PISN explosion.
The details of the components that construct the synthetic spectrum are summarized in
We have summarized in Appendix \ref{sec:app} and Figure \ref{fig:spectra_comp}.

Figure \ref{fig:spectra_day} shows the $\gamma$- and hard X-ray spectra at 150, 300, and 450 days after the explosion of a PISN with helium core mass $M_\mathrm{He}=130\,M_\odot$ located at a distance of 100 Mpc. 
A clear increase in the continuum-to-line flux ratio is observed as the supernova evolves from 150 to 450 days. Initially, at day 150, the regions containing $^{56}$Ni are optically thick to Compton scattering, leading to a reduced continuum-to-line flux ratio. 
From 150 to 450 days, the optical depth decreases, allowing a greater fraction of nuclear decay line photons to escape without scattering.
This effect, caused by photon losses due to Compton scattering processes, becomes evident 
when we focus on the spectra at energies above 200 keV in the bottom panel of Figure \ref{fig:spectra_day}.

\begin{figure}[htpb]
  \centering
  \includegraphics[width=0.45\textwidth]{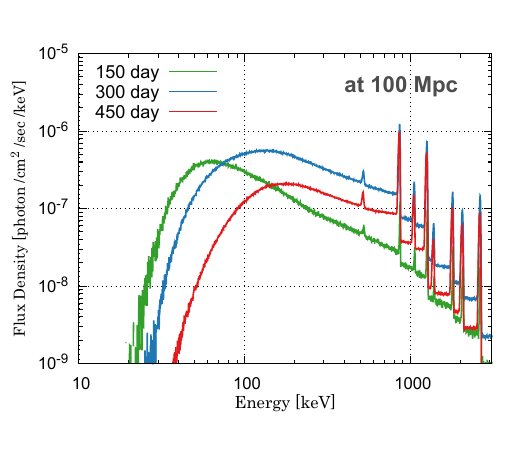}
  \includegraphics[width=0.45\textwidth]{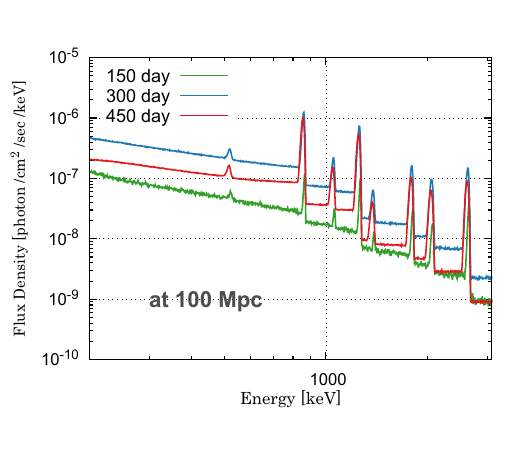}
  \caption{
    Spectra at 150 (green), 300 (blue), and 450 days (red) after the explosion of a PISN with helium core mass $M_\mathrm{He}=130\,M_\odot$ located at a distance of 100 Mpc. 
    }
\label{fig:spectra_day}
\end{figure}

Figure \ref{fig:spectra_model} presents a comparison of the $\gamma$- and hard X-ray spectra at 100 Mpc for different PISN models with helium core masses $M_\mathrm{He}=100\,M_\odot$ and $130\,M_\odot$. 
The blue $M_\mathrm{He}=130\,M_\odot$ model is the same as the 300 day model shown in blue in Figure \ref{fig:spectra_day}.
A notable feature is the broader line widths seen in the $130\,M_\odot$ model compared to the $100\,M\odot$ model, indicative of Doppler broadening effects. 
As shown in Figure \ref{fig:abundance}, the region enriched in $^{56}$Ni (defined as zones where $X(\mathrm{^{56}Ni})>0.1$) exhibits characteristic expansion velocities of $\lesssim5000$ km s$^{-1}$ in the $100\,M_\odot$ model, whereas in the $130\,M_\odot$ model, this region extends up to $\sim7000$ km s$^{-1}$. 
This higher velocity of the $^{56}$Ni-rich layers leads to stronger Doppler broadening of the emergent gamma-ray lines.

\begin{figure}[htpb]
\centering
  \includegraphics[width=0.45\textwidth]{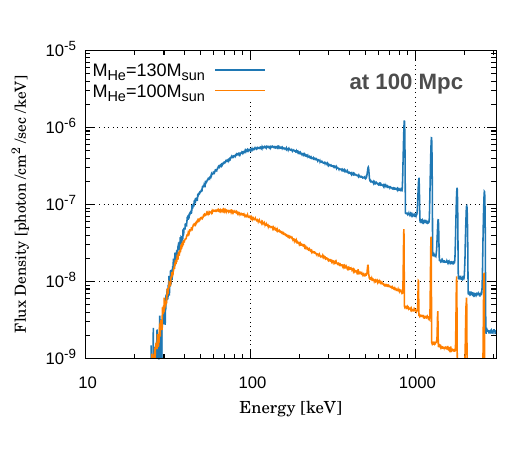}
  \includegraphics[width=0.45\textwidth]{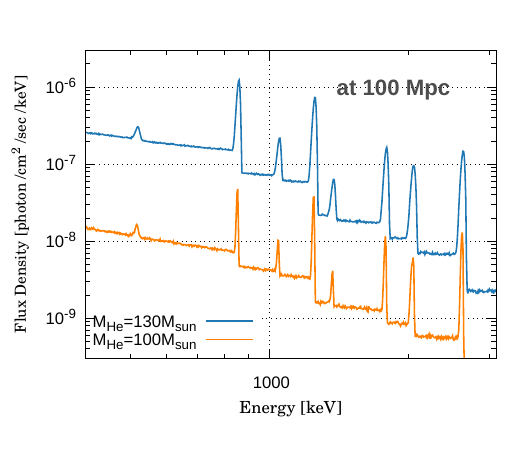}
    \caption{
    Spectra at 100 Mpc for different PISN models with helium core mass $M_\mathrm{He}=100\,M_\odot$ (orange) and $M_\mathrm{He}=130\,M_\odot$ (blue).
    }
\label{fig:spectra_model}
\end{figure}

Note that we neglect thermal Doppler broadening of the nuclear gamma-ray lines, as the effect is negligible compared to bulk expansion. 
For typical ejecta temperatures ( $T\sim10^9$ K), the thermal broadening is $\Delta E_\mathrm{th}\sim1$–$2$ keV at most, 
whereas bulk velocities ($v\sim 5,000$–$10,000\, \mathrm{km\,s^{-1}}$; see Fig. \ref{fig:abundance}) correspond to line widths $\gtrsim30$ keV.
Thus, the thermal contribution is subdominant by over an order of magnitude and can safely be neglected.

Both Figures \ref{fig:spectra_day} and \ref{fig:spectra_model} also exhibit a minor spectral feature around 511 keV, attributable to annihilation photons. 
We should note that our simulation included only para-positronium decay, but not ortho-positronium (see Appendix \ref{sec:app} and Figure \ref{fig:spectra_comp} in more detail), which could reduce the line flux and enhance the low-energy continuum \citep[e.g.,][]{2014Natur.512..406C}. 
A more complete treatment, such as that for recent Type Ia supernovae \citep[e.g.,][]{2025arXiv250522992D}, is left for future work.

\section{Light Curves}\label{sec:lc}

Figures \ref{fig:lc} show the light curves (in days after explosion) of the 847 and 1238 keV lines of $^{56}$Co decay at a distance of 100 Mpc for two PISN models with different helium core masses $M_\mathrm{He}=100\,M_\odot$ and $130\,M_\odot$.
For the 1238 keV line, the flux is computed by integrating a spectrum between 1150 and 1340 keV, and the 847 keV line flux was derived by integrating the spectrum between 800 and 920 keV. 
For the $M_\mathrm{He}=130\,M_\odot$ model, the peak emission is around day 300 after the explosion.

Although the total $^{56}$Ni mass in the $130\,M_\odot$ model is approximately three times that in the $100\,M_\odot$ model, the emergent gamma-ray flux is nearly an order of magnitude higher. 
This difference arises because, in the $100\,M_\odot$ model, the $^{56}$Ni is confined to the inner ejecta layers with lower expansion velocities (see Fig. \ref{fig:abundance}) and higher optical depths, reducing the effective escape fraction. 
In contrast, the $130\,M_\odot$ model distributes $^{56}$Ni over faster-moving, more extended regions, allowing more gamma rays to escape efficiently.

The 847 keV gamma-ray line flux of the $M_\mathrm{He}=130\,M_\odot$ model falls within the detection capabilities of the {\it INTEGRAL/SPI} instrument as well as next-generation $\gamma$-ray observatories, such as e-ASTROGAM. 
A simple estimation suggests a detection limit extending out to approximately $\sim300$ Mpc for the $M_\mathrm{He}=130\,M_\odot$ model. 
In contrast, the flux from the $M_\mathrm{He}=100\,M_\odot$ model remains significantly below detection thresholds, rendering its observation unrealistic at practical distances.

\begin{figure}[htpb]
\centering
  \includegraphics[width=0.45\textwidth]{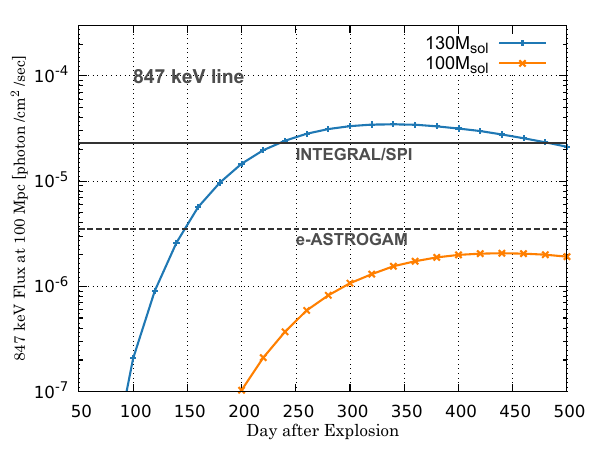}
\includegraphics[width=0.45\textwidth]{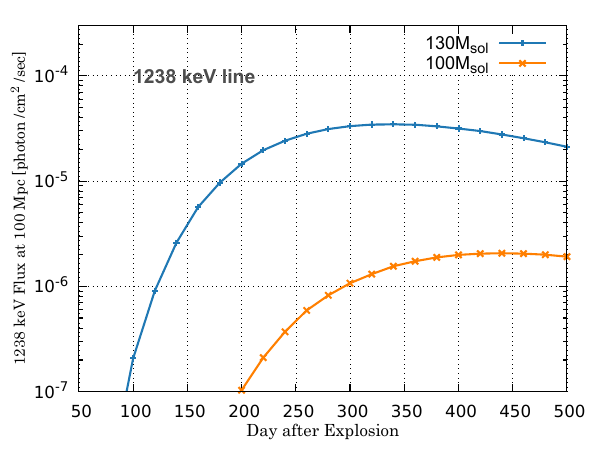}
    \caption{Light curves in photons cm$^{-2}$ s$^{-1}$ vs. time (in days after explosion) of the 847 and 1238 keV lines of $^{56}$Co decay at a distance of 100 Mpc for two PISN models with helium core masses $M_\mathrm{He}=100\,M_\odot$ (orange) and $M_\mathrm{He}=130\,M_\odot$ (blue).  
    }
\label{fig:lc}
\end{figure}

Additionally, Figure \ref{fig:lcx} shows the light curves of hard X-rays in the 45–105 keV range. These plots are particularly relevant to \textit{NuSTAR}.
The continuum component in the hard Xray range shifts toward higher energies over time, as can be seen in Figure \ref{fig:spectra_day}, and the light curves of hard X-rays reach their peak near 200 days for $M_\mathrm{He}=130\,M_\odot$ case and 300 days for $M_\mathrm{He}=100\,M_\odot$ case.

\begin{figure}[htpb]
\centering
  \includegraphics[width=0.45\textwidth]{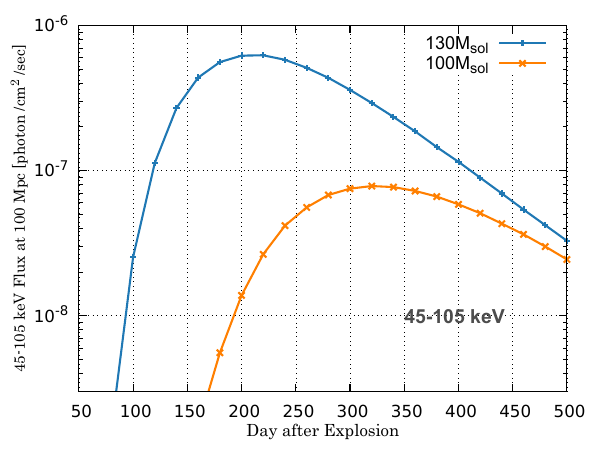}
    \caption{Light curves in the 45–105 keV range vs. time (in days after explosion) at a distance of 100 Mpc for two PISN models with helium core masses $M_\mathrm{He}=100\,M_\odot$ (orange) and $M_\mathrm{He}=130\,M_\odot$ (blue).  
    }
\label{fig:lcx}
\end{figure}

\section{Event rates of Pair-Instability Supernovae}\label{sec:discus}

Event rates are crucial for evaluating the detectability of PISNe within a 300 Mpc distance. 
However, their actual event rate remains uncertain due to the absence of confirmed observational detections of PISNe. 
Here, we estimate its value with reference to \cite{2022A&A...666A.157M}.

Assuming a Salpeter initial mass function \citep[][]{1955ApJ...121..161S} and a PISN progenitor mass range of $\sim$ 150–300 $M_\odot$ at the zero-age main sequence (ZAMS) stage \citep[][]{2002ApJ...567..532H}, we can estimate that event rates of PISN would be $\sim$1\% of core-collapse SNe ($M_\mathrm{ZAMS}\approx8–25\,M_\odot$). 
With core-collapse SN rates estimated to be between $10^5$–$10^6$ Gpc$^{-3}$ yr$^{-1}$ \citep[][]{2015ApJ...813...93S}, the corresponding predicted PISN rates within $\sim 300$Mpc would be roughly $10$ events yr$^{-1}$. 
Note that our hydrodynamical models represent the final pre-collapse masses of evolved He-core stars, after accounting for mass loss during stellar evolution.
As a rough correspondence, the initial ZAMS mass and the final core mass are related approximately as $M_\mathrm{core,f}\approx0.45M_\mathrm{ZAMS}$ \citep[][]{2014MNRAS.438.3119Y}, 
though this relationship, and in turn the mass range of PISNs, may vary due to mass loss processes such as stellar winds \citep[e.g.,][]{2019ApJ...887...53F,2025arXiv250510206S}.

However, observational constraints currently indicate that the PISN event rate must be lower than that of the SLSNe \citep{2021ApJ...908..249M}. 
Surveys have estimated the SLSN event rate to be about 30 Gpc$^{-3}$ yr$^{-1}$ at redshift $z \sim 0.2$ \citep[][]{2013MNRAS.431..912Q}. 
Given that the current number of PISN candidates is minimal compared to the roughly 100–200 confirmed SLSNe \citep[][]{2021ApJ...907L...9N,2022MNRAS.516.1072C}, the fraction of PISN relative to SLSN events is likely to be of the order of a few percent, leading the PISN event rates within $\sim 300$ Mpc to be approximately 0.01–0.1 events per yr$^{-1}$. 

Several supernovae have been proposed as potential candidates for PISNe, including SN 2007bi \citep[$ z = 0.13$;][]{2009Natur.462..624G}, OGLE14-073 \citep[$z = 0.12$;][]{2018MNRAS.479.3106K}, SN 2213-1745 \citep[$z = 2.045$;][]{2012Natur.491..228C}, SN 1000+0216 \citep[$z= 3.90$;][]{2012Natur.491..228C}, and SN 2018ibb \citep[$z=0.17$;][]{2024A&A...683A.223S}. 
These observational distances highlight that PISNe events are likely to be rare within a few hundred Mpc, but we can expect to observe at least one event over the course of a decade of observations.

\section{Conclusions}\label{sec:concle}

In this study, we have presented Monte Carlo radiation transport simulations of gamma-ray and hard X-ray emissions from pair-instability supernovae (PISNe), focusing on the radioactive decay chain of $\mathrm{^{56}Ni}\to\mathrm{^{56}Co}\to\mathrm{^{56}Fe}$. 
Despite the spherical symmetry assumed for the PISN ejecta structure, our Monte Carlo scheme fully treats the three-dimensional trajectories and interactions of photons.
Using detailed PISN models with helium core masses of $100\,M_\odot$ and $130\,M_\odot$, 
we have quantified the spectral evolution and line emissions, emphasizing detectability by current and future observatories.
Our key findings include: 
\begin{itemize}
    \item Based on our simulations, the maximum detection distance for the gamma-ray line emission, at 847 and 1238 keV from $\mathrm{^{56}Co}$ decay, is estimated to be up to $\sim300$ Mpc, assuming the single-epoch 5$\sigma$ sensitivity of next-generation MeV gamma-ray observatories, such as e-ASTROGRAM.
    \item The $100\,M_\odot$ helium-core mass model yields significantly lower gamma-ray fluxes, likely remaining below practical detection limits, highlighting the sensitivity of observational prospects to progenitor mass.
    \item Hard X-ray continuum emission, primarily due to Compton scattering of gamma-ray photons, demonstrates significant time evolution. Initially suppressed by high optical depth, the continuum flux increases notably as the ejecta expands and becomes optically thin, peaking at 200 days post-explosion for $M_\mathrm{He}=130\,M_\odot$ model and at 300 days for $M_\mathrm{He}=100\,M_\odot$ model.
    \item Event rate estimations for PISNe within 300 Mpc remain uncertain due to a lack of definitive observational confirmations. However, considering theoretical predictions and current observational constraints, we estimate the event rate to be approximately 0.1 per year. Although rare, dedicated gamma-ray observations over a decade-scale timespan could realistically capture these events.
\end{itemize}

We emphasize that, while our Monte Carlo radiotransport calculations fully track the trajectories and interactions of photons in three dimensions, our model assumes spherical symmetry in the structure of the PISN ejecta and in the spatial distribution of the photon sources corresponding to the $^{56}$Ni-rich regions. 
As discussed in Section \ref{subsec:bm}, multidimensional effects such as Rayleigh–Taylor instabilities are expected to cause only minor deviations from spherical symmetry \citep[e.g.,][]{2017A&A...599L...5G}, justifying this assumption.
More importantly, we should note that our current predictions are likely optimistic, as we have modeled progenitors without extended hydrogen envelopes. 
\cite{2019ApJ...882...22A} has shown that the presence of hydrogen-rich outer layers can significantly increase the optical depth and suppress the emergent gamma-ray flux, making our escape fraction estimates effectively an upper bound.

In conclusion, our detailed predictions emphasize that gamma-ray and hard X-ray observations represent a crucial avenue for the confirmation and detailed study of PISNe. 
Given the sensitivity enhancements provided by upcoming MeV gamma-ray observatories, PISNe are strongly anticipated as key target objects for future observational campaigns. 
This estimate is grounded on current instrument concepts with comparable sensitivity, including AMEGO-X and COSI, and highlights the observational potential of future MeV missions targeting radioactive decay signatures in PISNe.
The observational signatures outlined in this paper provide a clear target for upcoming gamma-ray observatories, potentially leading to a groundbreaking identification of this long-predicted, yet elusive class of stellar explosions.

\begin{acknowledgments}
We sincerely thank the anonymous referee for their thoughtful and constructive feedback, which has significantly improved the quality of this manuscript.
The work has been supported by Japan Society for the Promotion of Science (JSPS) 21K13964 (RS).
\end{acknowledgments}

\begin{contribution}
All authors contributed equally to this work.
\end{contribution}


\software{{\tt MESA} version {\tt 24.08.01} \citep{2011ApJS..192....3P,2013ApJS..208....4P,2015ApJS..220...15P,2018ApJS..234...34P,2019ApJS..243...10P,2023ApJS..265...15J}}

\restartappendixnumbering
\appendix
\section{ Contributions of various components to the model spectrum}\label{sec:app}
The contributions of various components to the total spectrum are shown in Appendix Fig. \ref{fig:spectra_comp}. 
The most prominent are the lines at 847 and 1238 keV. 
These lines are broadened by the expansion velocity of the ejecta and escape without any interactions. 
The line shape is also modified by opacity effects because the $\gamma$-ray photons produced on the approaching side of the ejecta have a higher chance of reaching an observer. 
This effect causes the line shapes to be skewed towards the blue side. 
The scattered continuum associated with the most prominent lines extends down to energies $\sim$100 keV.
At lower energies, photoabsorption becomes dominant, and the flux drops.

One additional component not included in our calculation is the three-photon decay of ortho-positronium, which can contribute to the continuum emission below 511 keV. 
Before annihilation, the positrons can form a bound state with an electron known as positronium \citep[][]{2004IJMPA..19.3879K}.
Positronium can be formed in either singlet (para) or triplet (ortho) spin states \citep[][]{1949PhRv...75.1696O}, with the latter decaying into three photons that share the 1022 keV rest mass energy \citep[][]{1976ApJ...210..582C}. 
This process can produce a low-energy continuum component independent of Compton scattering, particularly at late times \citep[][]{2014Natur.512..406C}. 
Our current model assumes only two-photon annihilation and neglects ortho-positronium formation.

\begin{figure}[htpb]
\centering
  \includegraphics[width=0.55\textwidth]{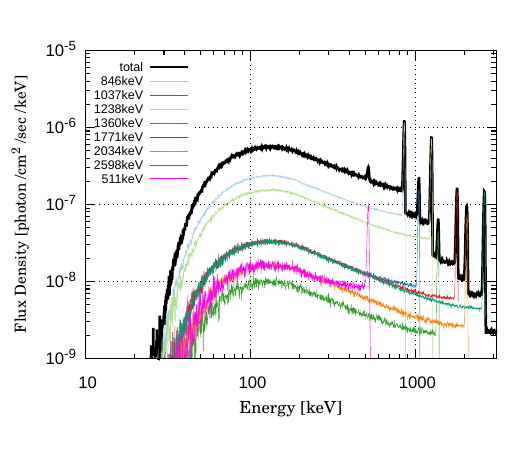}
    \caption{
    Contributions of various components to the model spectrum at 300 days 
    after the explosion of a PISN with helium core mass $M_\mathrm{He} = 130 M_\odot$ located at a distance of 100 Mpc.
    The lines are formed by $\gamma$-rays escaping the ejecta without interactions. 
    The low-energy tail of each line is due to Compton down-scattering of the photons because of the recoil effect. 
    The magenta line shows the contribution of the positronium annihilation. }
\label{fig:spectra_comp}
\end{figure}

\bibliography{ref}{}
\bibliographystyle{aasjournal}

\end{document}